\documentclass[preprint]{iacrtrans}

\usepackage{cite}
\usepackage[utf8]{inputenc}
\usepackage{amsmath}
\usepackage{graphicx}
\usepackage{subfigure}
\usepackage{appendix}
\usepackage{url}
\usepackage{diagbox}
\usepackage{hyperref}

\author[Minhui Jin et al.]{{Minhui Jin, Mengce Zheng, Honggang Hu, Nenghai Yu}}
\institute{Key Laboratory of Electromagnetic Space Information, CAS\\ University of Science and Technology of China, Hefei 230026, China \email{{jmh123,mczheng}@mail.ustc.edu.cn}}
\title[An Enhanced Convolutional Neural Network in Side-Channel Attacks]{An Enhanced Convolutional Neural Network in Side-Channel Attacks and Its Visualization}

\begin{document}

\maketitle


\keywords[Side-Channel Attack, Convolutional Neural Network, Attention Mechanism, Visualization Method]{Side-Channel Attack \and Convolutional Neural Network \and Attention Mechanism \and Visualization Method}

\begin{abstract}
In recent years, the convolutional neural networks (CNNs) have received a lot of interest in the side-channel community. The previous work has shown that CNNs have the potential of breaking the cryptographic algorithm protected with masking or desynchronization. Before, several CNN models have been exploited, reaching the same or even better level of performance compared to the traditional side-channel attack (SCA). In this paper, we investigate the architecture of Residual Network and build a new CNN model called attention network. To enhance the power of the attention network, we introduce an attention mechanism - Convolutional Block Attention Module (CBAM) and incorporate CBAM into the CNN architecture. CBAM points out the informative points of the input traces and makes the attention network focus on the relevant leakages of the measurements. It is able to improve the performance of the CNNs. Because the irrelevant points will introduce the extra noises and cause a worse performance of attacks.
We compare our attention network with the one designed for the masking AES implementation called ASCAD network in this paper. We show that the attention network has a better performance than the ASCAD network. Finally, a new visualization method, named Class Gradient Visualization (CGV) is proposed to recognize which points of the input traces have a positive influence on the predicted result of the neural networks. In another aspect, it can explain why the attention network is superior to the ASCAD network. We validate the attention network through extensive experiments on four public datasets and demonstrate that the attention network is efficient in different AES implementations.
\end{abstract}

\section{Introduction}
Side-channel attack (SCA) was first proposed on several cryptographic systems by Paul Kocher in 1996 \cite{kocher1996timing}. Since then, it has aroused great interest in the security community. SCA is a class of cryptanalytic attacks but different from traditional cryptanalysis. It exploits the physical properties such as timing, power consumption \cite{kocher1999differential}, electromagnetic (EM) emanation \cite{agrawal2002side} and even sound \cite{genkin2017acoustic}. In the side-channel community, SCA is divided into two categories: profiled attack and non-profiled attack. In the profiled attack, attackers have a copy of the target device. They can fully control of the input plaintexts and secret key of the replicated device and gather sufficient measurements (or traces). Then they build a model to describe the physical characteristics. Attackers subsequently utilize the measurements captured from the target device and perform the key-recovery process. The  profiled attack includes Template Attacks (TA) \cite{chari2002template} and Stochastic models \cite{schindler2005stochastic}. For the non-profiled attack, there is a weaker assumption. Attackers only have a target device, and they do not know the secret key about the encryption algorithm except the plaintexts and ciphertexts. They extract abundant traces from the target device and then exploit the statistical analysis technique to recover the secret key. The non-profiled attack includes Differential Power Analysis (DPA), Correlation Power Analysis (CPA) \cite{brier2004correlation} and Mutual Information Analysis (MIA) \cite{gierlichs2008mutual}. The attack methods mentioned above are called traditional SCA in this paper. In practical attacks,  pre-processing is required in traditional SCA like reducing the noise of the traces or selecting Points of Interests (PoIs) before recovering the secret key. Meanwhile, with the development of countermeasures such as desynchronization \cite{cagli2017convolutional} and masking \cite{maghrebi2016breaking}, traditional SCA becomes more difficult.

Convolutional neural networks (CNNs) have recently been introduced as a new alternative method to SCA. Researchers found that the profiled attack is similar to the classification problem in CNNs. For the profiled attack, attackers build a model to describe the features in the profiling stage which corresponds to the modeling stage in CNNs. Then attackers recover the secret key in the attacking stage which corresponds to the classifying stage in CNNs.
Thereby, one aspect of SCA is to exploit CNNs to recover the secret key of embedded devices. Before, CNNs have led to an impressive performance in many different fields where massive data are available, such as visual recognition tasks, natural language processing, medical data analysis, etc. In the field of SCA, CNNs have become one of the most powerful attack methods. In certain situations, they are even better than traditional SCA. For instance, some recent papers have implied that CNNs are at least as efficient as TA and under some circumstance, even better than TA\cite{zaid2020methodology,kim2019make}, while TA is considered to be the most powerful attack from an information-theoretic point of view. Compared to the traditional SCA, due to the translation invariance and some other natural characteristics, CNNs are robust to the most common countermeasures like desynchronization or masking \cite{cagli2017convolutional,zaid2020methodology,kim2019make,benadjila2018study}. Another advantage is that CNNs are efficient to deal with the multi-classification problems to implement end-to-end classifications. Moreover, CNNs can automatically extract features, so pre-processing is not needed.

In SCA, there exist the environment noises in the measurements. Traditional SCA and CNNs are all expected to focus on the informative points of the traces as far as possible. Because the irrelevant points will introduce extra noises and cause a worse performance of attacks. For the traditional SCA, it uses Principal Components Analysis (PCA) or Kernel Discriminant Analysis (KDA) to select the PoIs where most information is contained. But for the profiled SCA based on CNNs, it has not been considered yet. So in this paper, we propose a new CNN model. In the model, we introduce an attention mechanism - Convolutional Block Attention Module. CBAM can make the model focus on the important points and suppressing unnecessary points. Then we proposed a visualization technique called Class Gradient Visualization to verify the effectiveness of the new CNN model. 

\subsection{Related Work}
We introduce the application of neural networks in SCA in previous work. One of the first applications of neural networks in SCA was presented in \cite{yang2011back}. Subsequently, the applications of multiple-layer perceptron (MLP) in SCA have been proposed in \cite{martinasek2013optimization,martinasek2015power}. In \cite{cagli2017convolutional},  Cagli et al. proposed a data augmentation method and noted that CNN can deal with the traces misalignment. \cite{kim2019make} addressed how to improve the performance of the neural network by adding the artificial noise to input traces. \cite{zhou2019deep} showed that the synchronization method is still useful to CNNs. \cite{yang2018convolutional} exploited the time-frequency patterns of side-channel traces and perform a successful attack based on CNNs. In \cite{hettwer2018profiled}, Hettwer et al. used the domain knowledge to recover the secret key without any assumption about the leakage.

Another topic is to explain how the trained CNNs work and use the trained models to estimate the informative points of the input trace.
\cite{masure2019gradient} proposed a gradient visualization method to localize PoIs based on a successful trained neural network. It depends on the gradient of the loss function concerning the components of the input trace. In \cite{timon2019non}, Timon et al. introduced a Sensitivity Analysis to reveal the secret key and PoIs of the input trace. The method is similar to the gradient visualization in \cite{masure2019gradient}. In \cite{zaid2020methodology},  Zaid et al. explained the role of hyperparameters using some specific visualization techniques including Weight Visualization, Gradient Visualization, and Heatmap. \cite{hettwer2019deep} investigated three attribution methods especially the Layer-wise Relevance Propagation (LRP) to reveal the PoIs of the input traces. \cite{perin2019neural} evaluated the neural network using a  backward propagation path method. It can verify what the neural network learned from the side-channel traces.
\subsection{Our Contributions}
In this paper, our main contributions are summarized as follows.
\begin{enumerate}
\item We investigate the architecture of Residual Network and propose a new CNN model. To reduce the noises introduced by irrelevant points, we enhance the attention network by introducing an attention mechanism --- Convolutional Block Attention Module. CBAM can make the CNN models focus on the informative part.
\item We propose a new visualization method called Class Gradient Visualization (CGV) to recognize what the networks focus on in the training process and verify the effectiveness of the attention network. 
\item We validate the attention network through extensive experiments on four publicly available datasets, and the results show that the new CNN model is efficient in different AES implementations.
\end{enumerate}
\subsection{Organizations}
The rest of the paper is organized as follows. In \autoref{Section 2}, we introduce the notations of SCA. Then introduce the profiled attack and evaluation metrics. At last, we introduce an attention mechanism. \autoref{Section 3} introduces the architecture of the new CNN model and the application of attention mechanism in the new network model. We perform the experiments to evaluate the performance of the new CNN model in \autoref{Section 4}. In \autoref{Section 5}, we proposed a new visualization method and implement the visualization of network models. Finally, a conclusion is present in \autoref{Section 6}.



\section{Preliminaries}\label{Section 2}
The first part of this section provides notations of the side-channel attack and the subsequent parts introduce the profiled attack and the metric which is used to evaluate attacking result. At last, we briefly introduce an attention mechanism --- Attention Module in Attention Network.

\subsection{Notations}
In this paper, we use the calligraphic letter $\mathcal{X}$ to denote sets and the corresponding upper-case letter $X$ to denote random variable.
In SCA, the random variable is constructed as $X \in \mathbb{R}^{1\times D}$, where $D$ denotes the dimension of the traces. The lower-case letter $\vec{x}$ denotes the realization of $X$. The $i$-th entry of a vector $\vec{x}$ is denoted by $\vec{x}[i]$ and the $i$-th observation of a random variable $X$ is denoted by $\vec{x}_i$. For the encryption algorithm, there is $Z = f(P, K)$, where $f$ denotes the cryptographic primitive and $Z$ is the target sensitive variable. $P$ denotes the public variable (e.g. plaintext or ciphertext) and $K$ is the secret key that the attacker aim to retrieve. We denote $k^*$ as the secret key of the cryptographic algorithm and $k$ is any possible key hypothesis.
\subsection{Profiled Attack}
For the profiled attack, there are two phases: profiling phase and attacking phase. During the profiling phase, attackers have a copy of the target device. They set the plaintexts and the secret key of the cryptographic algorithm in the replicated device and collect sufficient measurements $X$. For each measurement $\vec{x}$, attackers compute the sensitive variable $z$ with the known information of the cryptographic algorithm. Attackers aim to estimate the probability 
\begin{equation}
Pr[X = \vec{x}|Z=z]
\end{equation}
by generating a model $F$ : $\mathbb{R}^D \rightarrow \mathbb{R}^{|Z|}$ from the profiling set $\{(\vec{x}_i,z_i)\}_{i=1,\dots,N_p}$, where $N_p$ denotes the number of measurements. Then in the attacking phase, attackers randomly choose plaintext $p_i$ of the target device and collect some measurements. The attacking set is  $\{(\vec{x}_i,p_i)\}_{i=1,\dots, N_a}$, where $N_a$ is the number of measurements. The secret key $k^*$ is unknown but fixed. Aimed to recover the secret key, attackers utilize the attacking set and the model $F(\cdot)$ built in the profiling phase and compute the score vector $F(X)$ of the sensitive variable Z. Then they use the Maximum Likelihood strategy to compute the predicted probability of each key candidate. The key that has the maximum probability is considered as the recovered key $k$. when the secret key $k^*$ is exactly the recovered key $k$, the profiled attack is successful.

\subsection{Evaluation Metric} In SCA, the common metrics to assess the performance of models are the Success Rate (SR) and Rank \cite{benadjila2018study}. In this paper, we use the Rank to evaluate the performance of models. Given $N_a$ attacking measurements, the key guess vector is $\vec{g} = \{ g_1, g_2,\dots, g_{|\mathcal{K}|}\}$ in descending order of the predicted probability, where $|\mathcal{K}|$ denotes the size of the keyspace. $g_1$ is considered as the recovered key of the models. the key guess vector is calculated by the log form of maximum likelihood strategy, i.e.,
\begin{equation}
g_i = \sum_{j=1}^{N_a}\log(\hat{p}_{ij}),
\end{equation}
where $\hat{p}_{ij}$ denotes the predicted probability of i-th key candidates in the j-th attacking measurement. Rank denotes the average position of $k^*$ in $\vec{g}$. When Rank is equivalent to $0$, it represents that $k^*$ is equivalent to $g_1$ and implies this attack is successful. The aim of our attack is to use the minimum number of attacking traces to achieve a successful attack.

\subsection{Convolutional Block Attention Module}
Attention mechanisms are inspired by the human visual system. One does not attempt to concentrate on the whole scene immediately. Instead, humans selectively focus on some important parts and recognize the scene \cite{zeiler2014visualizing}.
In \cite{woo2018cbam}, Woo et al. proposed a Convolutional Block Attention Module (CBAM) for feed-forward convolutional neural networks. CBAM separately underlines features along two dimensions: channel and spatial. Channel attention focuses on the meaningful intermediate features, and spatial attention stress where is the informative part.

For channel attention module, the module independently uses a Global Average Pooling (GAP) operation and a Global Max Pooling (GMP) operation to extract the average features and max features of the input features. Then both features are sent to a shared layer composed of MLP which has a hidden layer and produces the channel attention maps. The maps are considered to be the weight of each feature. Finally, an element-wise multiplication is computed between the input feature and the channel attention maps.
In short, channel attention can be characterized by the following formula.
\begin{equation}
\text{M}_\text{c}(\text{F}) = \sigma(MLP(GAP(\text{F})) + MLP(GMP(\text{F}))) \otimes \text{F}
\end{equation}
where $\sigma$ denotes the sigmoid function and $\otimes$ denotes the element-wise multiplications. $\text{M}_\text{c}$ and $\text{F}$ denote the input and output of the channel attention module.

For spatial attention module, the module first applies an average-pooling operation (AP) and a max-pooling (MP) operation along the channel axis of input features and generate an average samples and a max samples. Then it concatenate these two samples and apply a convolutional layer to generate the spatial attention maps. This maps can be viewed as the weight of each time points. Finally, an element-wise multiplication is computed between the input features and the spatial attention maps. Spatial attention module is computed as:
\begin{equation}
\text{M}_\text{s}(\text{F}) = \sigma( \gamma([AP(\text{F}); MP(\text{F})])) \otimes \text{F}
\end{equation}
where $\sigma$ denotes the sigmoid function and $r$ denotes the convolutional layer. $\text{M}_\text{s}$ denotes the output of the spatial attention module.

\section{CNN Architecture} \label{Section 3}
In this section, we introduce the architecture of the new CNN model and the application of CBAM in the new network. CBAM makes the attention network focus on the informative points and suppresses irrelative points to improve the performance of CNNs. 
\subsection{Basic Architecture of the Enhanced Network}

\begin{figure}[h]
\centering
\includegraphics[scale=0.35]{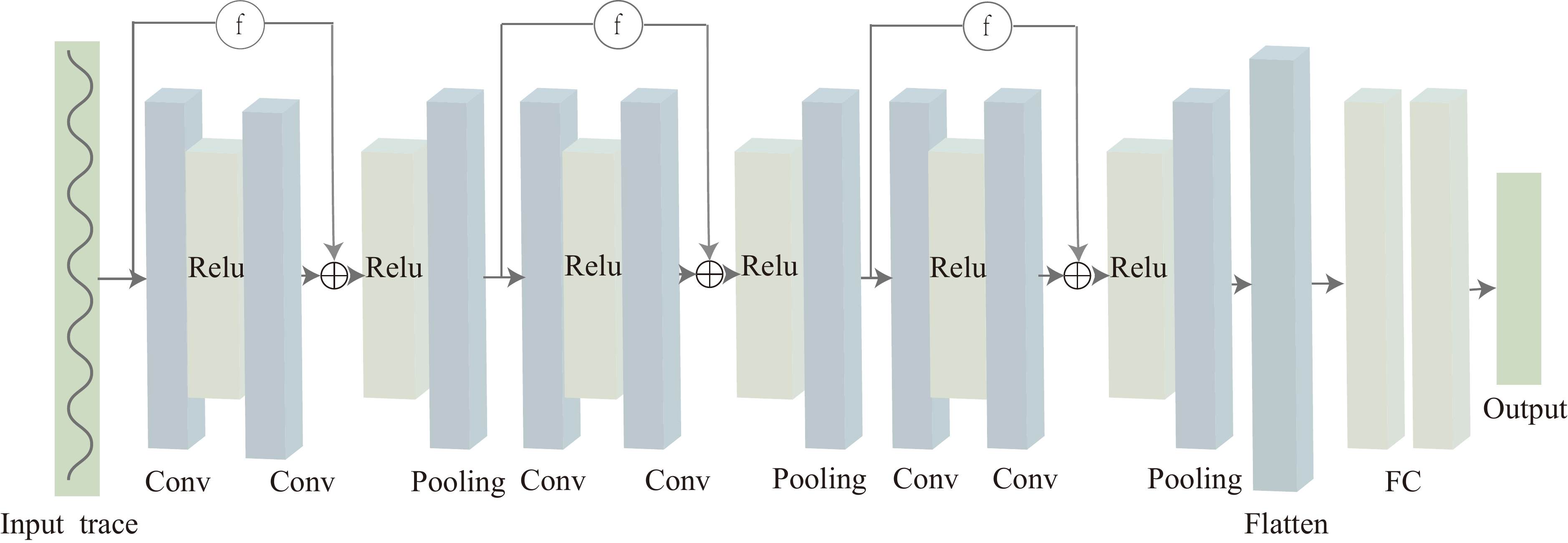}
\caption{The basic architecture of the attention network. As the architecture is shown, the input of the attention network is the whole traces. The output layer has 256 neurons which correspond to the 256 classes. We use 3 residual blocks and each residual block consists of 2 basic blocks. The function $f$ is a convolutional operation applied to the inputs of the basic blocks.}
\label{fig 1}
\end{figure}

With the development of CNNs, some classical architectures have been proposed such as VGG \cite{simonyan2014very} and Residual Networks (ResNets) \cite{he2016deep}. Some researchers have investigated the application of VGG architecture in SCA \cite{benadjila2018study,kim2019make} and indicated that the architecture performs well. In this work, we investigate another architecture - ResNets in SCA and propose a new CNN model called attention network. The basic structure of the attention network is shown in \autoref{fig 1}. It consists of several blocks with similar architecture. These blocks are called residual blocks. Each residual block is composed of several layers and a shortcut connection. In the attention network, these several layers are convolutional layers $\gamma$, pooling layers $\delta$, and activation functions $\sigma$. Each residual block has two basic blocks. The basic block is composed of convolutional layers and activation functions. The shortcut connection connects the input and output of the basic block. It can deal with the degradation problem. After stacking several residual blocks, the flatten layer and fully-connected layers $\lambda$ are adopted. Finally, a softmax layer $s$ is applied to generate the predicted probability of each class.

The structure of attention network can be characterized by the following formula:
\begin{equation}
s \circ [\lambda]^{n_1} \circ [\delta \circ \sigma \circ [\gamma(x) \circ \sigma]^{n_2} \oplus f(x)]^{n_3}
\end{equation}
where $n_1$, $n_2$, $n_3$ represent the number of fully-connection layers, basic blocks, and residual blocks. In attention network, $n_1$ and $n_2$ are all set to 2, while $n_3$ is up to the datasets. Shortcut connection is denoted as $\oplus$. Here $x$ is the input of the current basic block and $f(x)$ is the function of $x$. The funtion $f$ is to make the input and output of basic block have the same dimensions and is a convolutional operation in attention network. The basic hyperparameters in the attention network are set as follows:
\begin{itemize}
	\item For the convolutional layers, the size of the filter is set to 11 and the stride is set to 1. The number of the filter is various in different convolutional layers.
	\item For the pooling layers, we use average-pooling. The pooling stride is set to 2 and the pooling window is the same.
	\item The activation functions are all set to 'ReLU'.
\end{itemize}
The other hyperparameters are up to the experimental datasets and they will be shown later in the paper. The choices of hyperparameters are motivated by the fact that they provide good results based on our datasets.

\subsection{Enhanced CNN with Attention Module}
\begin{figure}[h]
	\centering
	\includegraphics[scale=0.45]{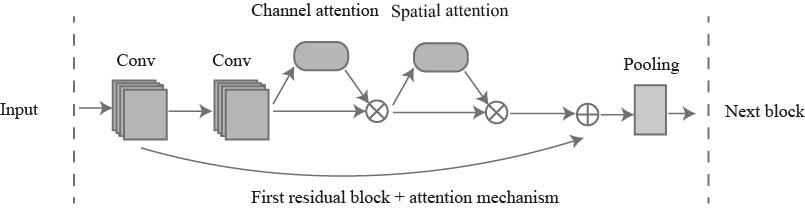}
	\caption{Attention mechanism in the attention network.}
	\label{fig 2}
\end{figure}

For CNNs, it can automatically extract the different features of the whole trace by different filters of the convolutional operation. But the unnecessary points of the input traces not only do not provide useful information, but also will introduce extra noises. So some features extracted by filters are still useless and worsen the performance of CNNs. To make the attention network focus on the informative points, we introduce an attention mechanism - Convolutional Block Attention Module and apply it to the attention network. CBAM can strengthen the representation of the features in the attention network. First, the channel attention module is used to assign weights to different features. By this module, attention network can find the important features. Besides, spatial attention is used to point out the locations of the features. So we use the spatial attention module to locate the important features after the channel attention module. As is shown in \autoref{fig 2}, the attention module is inserted in the first residual block and sequentially uses channel attention module and spatial attention module. By introducing the CBAM, the attention network can focus on the informative points of the traces and reduce the noises introduced by the unnecessary points.



\section{Performance of the Enhanced Network} \label{Section 4}
In this section, we apply our attention network on different publicly available datasets. These datasets will be introduced in the \autoref{Section 4.1} and the results will show in the \autoref{Section 4.2}. Our experiments are performed by the NVIDIA Graphics Processing Units (GPUs) and we implement the experiment with the open-source deep-learning library Keras and TensorFlow backend. 
\subsection{Datasets}\label{Section 4.1}
In our experiments, we use four public datasets. All of them are the implementations of the Advanced Encryption Standard (AES) algorithm. They are divided into two types: unprotected implementations and protected implementations. The unprotected implementations have two types, for instance, software implementation and hardware implementation. The countermeasures of protected implementations also have two types such as desynchronization and masking. In the following, we will introduce these four datasets.
\subsubsection{DPAcontest v4}
The cryptographic algorithm of DPAcontest v4 is a masking software implementation of AES-256, which is called AES-256 RSM (Rotating Sbox Masking) on an Atmel ATMega-163 smart-card. DPAcontest v4 is electromagnetic measurements where it uses the same secret key with different plaintexts. \cite{moradi2014detecting} has verified that the masking implementation exists the first-order leakage, so we turn the masking protected implementation into the unprotected scenario. The corresponding leakage model is changed to:
\begin{equation}
Y(k^*) = Sbox[P_i \oplus k^*] \oplus M
\end{equation}
where $M$ denotes the mask of each cryptographic operation and the value is known. $P$ corresponds to the plaintexts. We choose the first Sbox operation in the first round of cryptographic operation that $i$ is set to 1. 

For DPAcontest v4, the measurements consist of 10 000 traces, each trace with 1000 samples. We divide the dataset into two subsets such that 5000 traces for the training set and 5000 traces for the attacking set. For the network settings of DPAcontest v4, there are 3 residual blocks and the number of the filters of the convolutional layers in each residual block is set to \{128, 256, 512\}. The epoch is set to 60 and the batch-size is set to 200. The number of neurons of the hidden fully-connected layers is set to 1024. This public dataset is available at \url{http://www.dpacontest.org/v4/}.
\subsubsection{AES\_RD}
The cryptographic algorithm in AES\_RD is a protected software implementation of AES. It is one of the typical classes used in reality. The target device is a smartcard of 8-bit Atmel AVR microcontroller. The countermeasure of the algorithm applies random delay which is proposed in \cite{coron2009efficient}. Adding random delay in the algorithm causes the misalignment of the time samples and make the attack more difficult. For this dataset, the leakage model is defined as follows:
\begin{equation}
Y(k^*) = Sbox[P_i \oplus k^*]
\end{equation}
As same as DPAcontest v4, we choose the first Sbox operation in the first round of cryptographic operation that is $i=1$.

For AES\_RD, the measurements consist of 50 000 traces and each trace has 3500 time samples. In the experiment, we divide the dataset into two subsets such that 40 000 traces for training and 10 000 traces for attacking. For the network settings of AES\_RD, there are 5 residual blocks and the number of the filters of the convolutional layers in each residual block is set to \{64, 64, 128, 128, 256\}. The number of neurons of hidden fully-connected layers is set to 1024 and we add two dropout layers whose rates are all set to $0.2$. In the training process, the epoch is set to 101 and batch-size is set to 256.
This dataset is available at \url{https://github.com/ikizhvatov/randomdelays-traces}.
\subsubsection{AES\_HD}
The cryptographic algorithm in AES\_HD is a typical class of the hardware implementations of AES-128. The hardware implementation was written in VHDL and applied the parallel operation that it takes 11 clock cycles for each encryption. The hardware design was implemented on Xilinx Virtex-5 FPGA of a SASEBA GII evaluation board. AES\_HD is electromagnetic measurements that were measured using a high sensitivity near-field EM probe. The suitable and common leakage model of unprotected hardware implementations is to exploit the register writing in the last round. The leakage model is defined as follows:
 \begin{equation}
Y(C_{i_1}, C_{i_2}, k^*) = Sbox^{-1}[C_{i_1} \oplus k^*] \oplus C_{i_2}
\end{equation}
where $C_{i_1}$ and $C_{i_2}$ denote two ciphertexts, and the relation between $i_1$ and $i_2$ is given through the inverse ShiftRows operation of AES. we choose $i_1 = 12$ and $i_2 = 8$.

The measurements of AES\_HD is composed of 500 000 traces with 1250 samples of each trace. In our experiment, we only use 75 000 traces that 50 000 traces are used for training and 25 000 traces are used for attacking. The network settings of AES\_HD are the same as DPAcontest v4, except that the epoch is set to 75. AES\_HD is publicly available at \url{https://github.com/AESHD/AES_HD_Dataset}.

\subsubsection{ASCAD}
The final dataset is ASCAD and the cryptographic algorithm is a masking software protected implementation of AES-128 \cite{benadjila2018study}. The implementation is running over an 8-bit AVR architecture that is ATMega8515. ASCAD possesses first-order security that is robust to the first-order SCA. Measurements are measured using an electromagnetic probe and stored with the current version 5 of the Hierarchical Data Format (HDF5). The leakage model is defined as follows:
 \begin{equation}
Y(k^*) = Sbox[P_i \oplus k^*]
\end{equation}
The first and the second Sbox operations in the first round of the encryption process are unprotected, so we choose the third Sbox operation that $i=3$.

This dataset is composed of 60 000 traces with 700 features of each trace. There are 50 000 in the training dataset group and 10 000 in the test dataset group. So we use the training dataset group to train the CNN model and use the test dataset group to perform attack. The network settings for ASCAD are the same as DPAcontest v4, except that the epoch is set to 75 and the number of neurons of the hidden fully-connected layers is set to 4096. ASCAD dataset is available at \url{https://github.com/ANSSI-FR/ASCAD}.

\subsection{Experimental Results}\label{Section 4.2}
We will show the performance of the attention network on four public datasets and compare it with the ASCAD network proposed in \cite{benadjila2018study} and TA performed in \cite{chari2002template}. For our experiments, we compare these two networks with the Average Rank which repeats the attacking process 300 times. The test traces are randomly selected from the test dataset on each attack. The aim of using Average Rank is to reduce the effect of the sequence of test traces. We verify that the attention network have the same or even better performance compared to the ASCAD network and TA on different AES implementations.
\subsubsection{DPAcontest V4}
\begin{figure}[htb]
	\centering
	\includegraphics[scale=0.57]{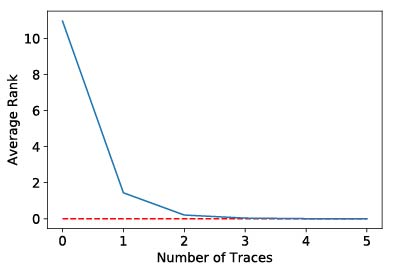}
	\caption{Average Rank in DPAcontest v4.}
	\label{Figure 3}
\end{figure}

The first dataset we try to attack is DPAcontest v4. In our experiments, DPAcontest v4 is the easiest dataset, because we apply the known masks to transform it into the unprotected implementation. The \autoref{Figure 3} shows that the attention network has an excellent result with several traces to attack successfully. The required traces are displayed in \autoref{Comparison on DPAcontest v4} that attention network demands 3 traces which is the same to the ASCAD network. TA needs 4 traces to recover the secret key. The result verified that when the AES is software unprotected, both CNN models perform excellently and have comparable performance to TA.

\begin{table}[htb]
	\centering
	\renewcommand{\arraystretch}{1.2}
	\caption{Comparison on DPAcontest v4}
	\label{Comparison on DPAcontest v4}
	\begin{tabular}{c|c|c|c}
		\multicolumn{1}{c|}{\textbf{}} & \multicolumn{1}{c|}{\textbf{\begin{tabular}[c]{@{}c@{}}ASCAD network\\ \cite{benadjila2018study}\end{tabular}}} & \multicolumn{1}{c|}{\textbf{\begin{tabular}[c]{@{}c@{}}Template attacks\\ \cite{kim2019make}\end{tabular}}} & \multicolumn{1}{c}{\textbf{Attention network}} \\ \hline
		\textbf{Required traces} & 3 & 4 & 3
	\end{tabular}	
\end{table}


\subsubsection{AES\_RD}
\begin{figure}[htb]
	\centering
	\includegraphics[scale=0.55]{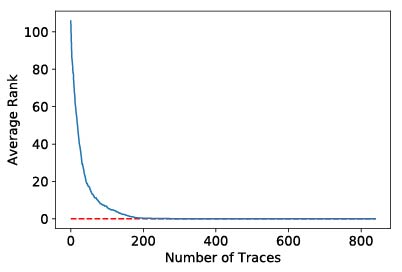}
	\caption{Average Rank in AES\_RD.}
	\label{fig 4}
\end{figure}

The algorithm of AES\_RD applies the random delay and makes it hard to align the PoIs. \autoref{fig 4} shows that we need more traces to implement a successful attack compared to DPAcontest v4.  The attention network needs hundreds of traces to make average rank equal to $0$. We also use the ASCAD network to attack the dataset and the result shows in \autoref{Comparison on AES_RD}. ASCAD network needs 247 traces to make the average rank less than $1$. But the attention network only needs 171 traces. The result shows both CNN models are robust to the desynchronization and attention network works better than the ASCAD network. For TA, it cannot implement an efficient attack even with 20 000 traces, which reveals that traditional SCA is hard to break the AES implementation with desynchronization.

\begin{table}[htb]
	\centering
	\renewcommand{\arraystretch}{1.2}
	\caption{Comparison on AES\_RD}
	\label{Comparison on AES_RD}
	\begin{tabular}{c|c|c|c}
		\multicolumn{1}{c|}{\textbf{}} & \multicolumn{1}{c|}{\textbf{\begin{tabular}[c]{@{}c@{}}ASCAD network\\ \cite{benadjila2018study}\end{tabular}}} & \multicolumn{1}{c|}{\textbf{\begin{tabular}[c]{@{}c@{}}Template attacks\\ \cite{kim2019make}\end{tabular}}} & \multicolumn{1}{c}{\textbf{Attention network}} \\ \hline
		\textbf{Required traces} & 247 & >20 000 & 171
	\end{tabular}	
\end{table}

\subsubsection{AES\_HD}

\begin{figure}[htb]
	\centering
	\includegraphics[scale=0.55]{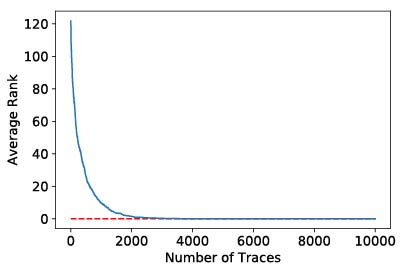}
	\caption{Average Rank in AES\_HD.}
	\label{fig 5}
\end{figure}

AES\_HD is an unprotected hardware AES implementation. It is difficult to attack because of a much higher level of environmental and algorithmic noise. \autoref{fig 5} shows that the attention network has a dramatic performance. \autoref{Comparison on AES_HD}
represents the required traces of the three models. The attention network only needs around 2100 traces to implement an efficient attack, while the ASCAD network and TA cannot recover the secret key even using the whole traces of the test dataset. It explicates that the attention network is much better than the ASCAD network and TA. We think it is because the attention network can focus on the regions of leakage and ignore the unnecessary points by using CBAM. It reduces the influence of the noises and improves performance.

\begin{table}[ht]
	\centering
	\renewcommand{\arraystretch}{1.2}
	\caption{Comparison on AES\_HD}
	\label{Comparison on AES_HD}
	\begin{tabular}{c|c|c|c}
		\multicolumn{1}{c|}{\textbf{}} & \multicolumn{1}{c|}{\textbf{\begin{tabular}[c]{@{}c@{}}ASCAD network\\ \cite{benadjila2018study}\end{tabular}}} & \multicolumn{1}{c|}{\textbf{\begin{tabular}[c]{@{}c@{}}Template attacks\\ \cite{kim2019make}\end{tabular}}} & \multicolumn{1}{c}{\textbf{Attention network}} \\ \hline
		\textbf{Required traces} & >25 000  & >25 000 & 2100
	\end{tabular}	
\end{table}

\subsubsection{ASCAD}
\begin{figure}[h]
	\centering
	\includegraphics[scale=0.55]{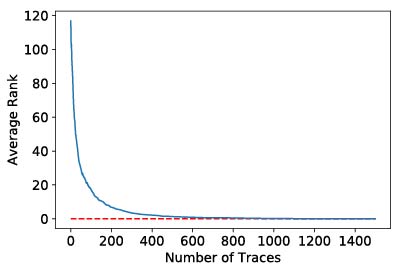}
	\caption{Average Rank in ASCAD.}
	\label{fig 6}
\end{figure}

The cryptographic algorithm of ASCAD applies masks to reduce the relevance between the measurements and sensitive values. \autoref{fig 6} shows that the attention network has efficient performance in ASCAD. \autoref{Comparison on ASCAD} represents the traces that are required to implement a successful attack. The attention network requires 550 traces, while the ASCAD network needs 770 traces. Though the ASCAD network is designed for the masking AES implementation, the attention network is still better than the ASCAD network. For traditional SCA, \cite{kim2019make} performed the template attack, but it did not retrieve the secret key. They used 500 traces and the Rank is still greater than 50. It shows the attention network is also better than TA.

\begin{table}[h]
	\centering
	\renewcommand{\arraystretch}{1.2}
	\caption{Comparison on ASCAD}
	\label{Comparison on ASCAD}
	\begin{tabular}{c|c|c|c}
		\multicolumn{1}{c|}{\textbf{}} & \multicolumn{1}{c|}{\textbf{\begin{tabular}[c]{@{}c@{}}ASCAD network\\ \cite{benadjila2018study}\end{tabular}}} & \multicolumn{1}{c|}{\textbf{\begin{tabular}[c]{@{}c@{}}Template attacks\\ \cite{kim2019make}\end{tabular}}} & \multicolumn{1}{c}{\textbf{Attention network}} \\ \hline
		\textbf{Required traces} & 769  & >500 & 552
	\end{tabular}	
\end{table}

\section{Network Visualization}\label{Section 5}
We first introduce the principle of a new visualization method --- Class Gradient Visualization. Then we use the visualization method to generate the visualization weight of each point in the traces and evaluate the effectiveness of the attention network and ASCAD network.
\subsection{Class Gradient Visualization}
Though CNNs have a significant performance in visual tasks, natural language processing, and even side-channel attack, it is still a 'black' box operation. In the field of computer vision, in order to understand how CNNs work, researchers have investigated the visualization technique. (GAP)\cite{lin2013network} is directly followed by the softmax layer. 
\cite{selvaraju2017grad} proposed the Gradient-weighted Class Activation Mapping (Grad-CAM). Grad-CAM uses gradients to locate the important regions in the image. It computes the gradients of the features of the last convolutional operations to the predicted class. But Grad-CAM is not suited for the attention network. Because it is only suited to the network using Global Average Pooling (GAP) after the last convolutional layer. In our network, we exploit the average pooling and flatten operation instead of GAP. Besides, Grad-CAM is used to visualize the regions of images which is 2-dimensions. While the measurements of SCA are always 1-dimension. Therefore based on Grad-CAM, we proposed a new visualization method in the side-channel context, called Class Gradient Visualization (CGV). 

The principle of the CGV is that the convolutional operation with the average pooling operation of the CNNs is viewed as the process of extracting features. Thus the effect of the feature maps of the last pooling operation on the predicted result can reflect which areas of input traces are important and positive to the final predictive class. In the CGV method, we compute the gradient of the features after the final pooling operation with respect to the class score.

For the predicted class $c$, $y^c$ denotes the class score that is before the softmax function. $A$ represents the feature matrix after the final pooling operation such that $A \in \mathbb{R}^{D \times V}$. Here $D$ denotes the number of samples and $V$ denotes the number of features. $x_i^j$ is an element of matrix $A$ where $i \in \mathbb{R}^V$ and $j \in \mathbb{R}^D$. We first compute the gradient of feature matrix $A$ with respect to the class score $y^c$ and obtain the weights matrix $W$ where $W \in \mathbb{R}^{D \times V}$. $\alpha_i^j$ denotes the weight of the j-th sample of the i-th feature:
\begin{equation}
\alpha_i^j = \frac{\partial y^c}{\partial x_i^j}
\end{equation}
Then we compute a weight map $W^c_{CGV}$ for class $c$ and each element of it is computed as follows:
\begin{equation}
W_{CGV}^c = ReLU(\sum_{i}^{V} x_i \odot \alpha_i)
\end{equation}
where $x_i \odot \alpha_i = (x_i^0\alpha_i^0, x_i^1\alpha_i^1, \dots, x_i^D\alpha_i^D)$. The larger value of the weight map shows that the corresponding time sample has a more important influence on the predicted class. In our experiment, we only explore the positive influence of the samples to the predicted class, so we use the 'ReLU' activation function to keep positive effective samples and discard the negative effective samples. Using the CGV visualization method, we get a coarse weight visualization of the same size as the pooling feature maps. Finally, we expand the size of the coarse weight visualization to the original size of the input traces. So we roughly estimate which regions of the input have more influence on the final predicted results.

\subsection{Visualization Results}
We visualize which components of the input traces have a positive influence on the final prediction by the CGV method in the trained networks. For the neural network, not all the points of the input traces are important. Irrelevant points will increase the noises and have negative effects on the prediction. Thus the CNN models that take more attention to the significant parts of the traces and take less attention to the unimportant parts will have a better performance. ASCAD network and attention network are the target models.

\begin{figure}[hb]
	\centering
	\subfigure[]{
		\label{fig 7.c}
		\includegraphics[scale=0.45]{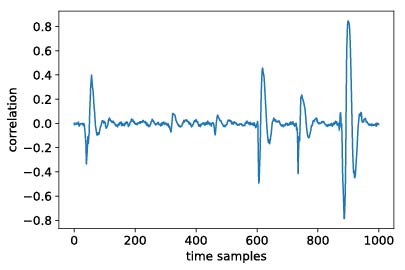}
	}

	\subfigure[]{
		\label{fig 7.a}
		\includegraphics[scale=0.45]{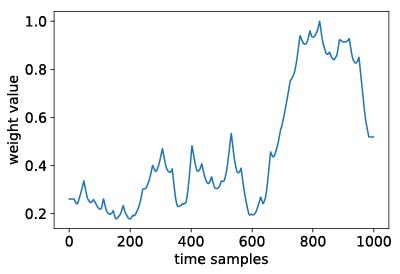}
	}
	\subfigure[]{
		\label{fig 7.b}
		\includegraphics[scale=0.45]{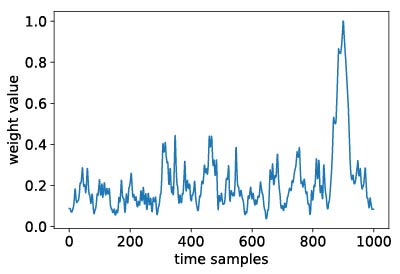}
	}
	\caption{(a) CPA analysis in DPAcontest v4; (b) Weight visualization of the ASCAD network in DPAcontest v4; (c) Weight visualization of the attention network in DPAcontest v4.}
	\label{fig 7}
\end{figure}

\begin{figure}[ht]
	\centering
	\subfigure[]{
		\label{fig 8.c}
		\includegraphics[scale=0.45]{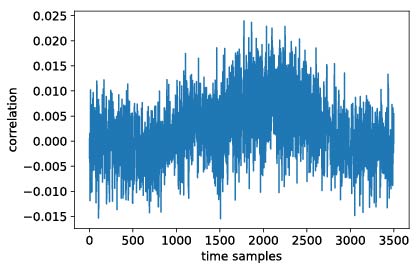}
	}

	\subfigure[]{
		\label{fig 8.a}
		\includegraphics[scale=0.45]{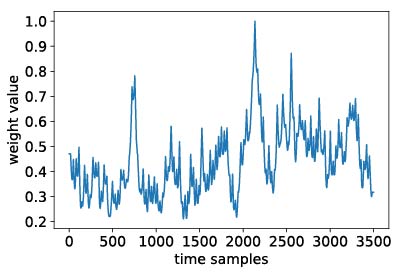}
	}
	\subfigure[]{
		\label{fig 8.b}
		\includegraphics[scale=0.45]{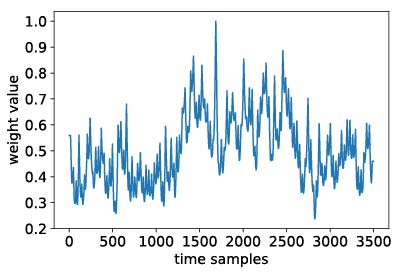}
	}
	\caption{(a) CPA analysis in AES\_RD; (b) Weight visualization of the ASCAD network in AES\_RD; (c) Weight visualization of the attention network in AES\_RD.}
	\label{fig 8}
\end{figure}

\begin{figure}[H]
	\centering
	\subfigure[]{
		\label{fig 9.c}
		\includegraphics[scale=0.45]{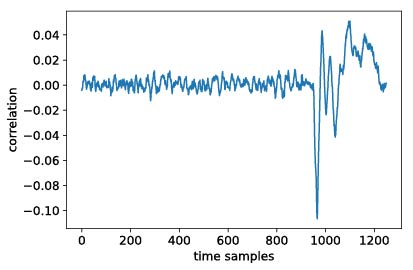}
	}

	\subfigure[]{
		\label{fig 9.a}
		\includegraphics[scale=0.45]{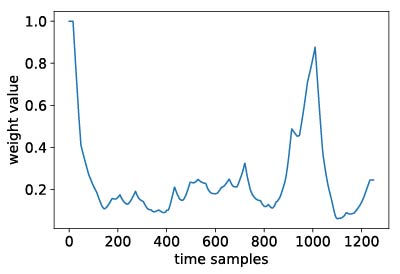}
	}
	\subfigure[]{
		\label{fig 9.b}
		\includegraphics[scale=0.45]{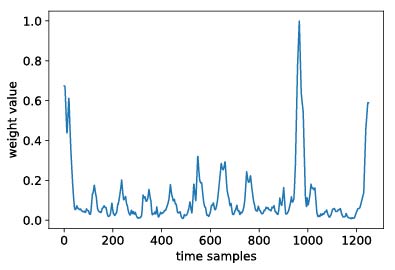}
	}
	\caption{(a) CPA analysis in AES\_HD; (b) Weight visualization of the ASCAD network in AES\_HD; (c) Weight visualization of the attention network in AES\_HD.}
	\label{fig 9}
\end{figure}

We utilize plaintexts and the secret key of cryptographic algorithms to exploit CPA analysis and find the leakages of the traces. \autoref{fig 7} shows the CPA analysis of DPAcontest v4 and the extended weight visualizations of the ASCAD network and the attention network. For the attention network, the visualization weight implies that the part of the largest weight corresponds to the part of the highest correlation coefficient in the CPA analysis. while the ASCAD network not only attends to the informative part but also learns the unimportant part around the informative area.
For AES\_RD, though it applies random delay, CPA analysis still finds the region of leakages. The weight visualization shows that the ASCAD network mainly focuses on the region between 2000 and 3500 in the input traces. But CPA analysis shows the samples between 3000 and 3500 have no apparent leakages. The region that the attention network focuses on is consistent in the region of leakages in CPA analysis (see \autoref{fig 8}).
As for AES\_HD, though measurements have a lot of noise, we still find the main leakages in CPA analysis. The region of leakages is between 980 and 1000. ASCAD network attends to the information of this area but still learns the other area which is the beginning of the traces. But the attention network mainly focuses on the area of leakages in the CPA analysis(see \autoref{fig 9}).

\begin{figure}[h]
	\centering
	\subfigure[]{
		\label{fig 10.c}
		\includegraphics[scale=0.45]{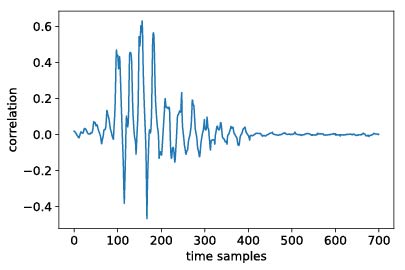}
	}
	\subfigure[]{
		\label{fig 10.d}
		\includegraphics[scale=0.45]{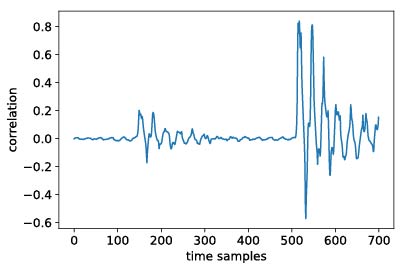}
	}
	\subfigure[]{
		\label{fig 10.a}
		\includegraphics[scale=0.45]{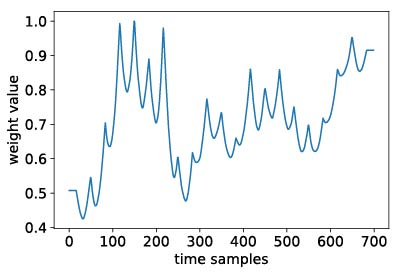}
	}
	\subfigure[]{
		\label{fig 10.b}
		\includegraphics[scale=0.45]{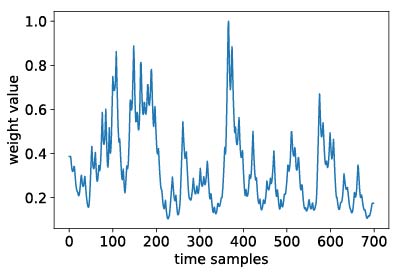}
	}
	\caption{(a) CPA analysis of masks in ASCAD; (b) CPA analysis of outputs of masked Sbox in ASCAD; (c) Weight visualization of the ASCAD network in ASCAD; (d) Weight visualization of the attention network in ASCAD.}
	\label{fig 10}
\end{figure}

For ASCAD, due to the existence of masks, the visualization weight of ASCAD is not apparent compared to the other three datasets. From the result of visualization (see \autoref{fig 10}), we find that the ASCAD network almost learns the whole trace, while the attention network mainly focuses on three regions. The CPA analysis of masks and outputs of masked Sbox shows that the samples between 50 and 400 leak the information of masks corresponding to the first region between 70 and 200 and the second region between 350 and 400. The third region around the 600 samples corresponds to the leakage information of masked Sbox output. By the CGV visualization method, It implies that the attention network takes more attention to the leakages compared to the ASCAD network. Especially for AES\_HD, attending to the leakages instead of other unnecessary points can improve the attacking performance a lot, because it reduces the effect of the noises.

\section{Conclusions} \label{Section 6}
In this paper, we investigate the architecture of ResNets in SCA and build a new CNN model. Besides, we introduce an attention mechanism --- Convolutional Block Attention Module (CBAM) and insert the module to the new CNN model. CBAM is to make the network attend to the informative part of the input traces and suppress the unnecessary points. Based on it, the new network can reduce the influence of the noise introduced by the irrelevant points and improve the performance.
Then we use four publicly available datasets to analyze the performance of the attention network and compare it to the ASCAD network. The experiments show that the attention network has a better performance than the ASCAD network in different AES implementations. We also compare the attention network to the traditional SCA such as TA, and the results show the performance of the attention network is extremely better than TA. 

Besides, we propose a new visualization method called Class Gradient Visualization. For a trained model, applying this method on the training traces can understand what the model focuses on and observe which points of the input traces have a positive influence on the final prediction. We use this visualization method to attention network and ASCAD network. The results show that the attention network takes more attention to the important leakage compared to the ASCAD network. In other respects, it can explain why the attention network works better than the ASCAD network. In this paper, we only exploit the CBAM and in future work, we will explore other attention mechanisms and compare the effectiveness of different attention mechanisms.


\bibliographystyle{alpha}
\bibliography{bibfile}




%



\end{document}